\documentclass[prl,twocolumn,amsmath,amssymb,nofootinbib]{revtex4}
\usepackage{graphicx} 
\usepackage{amsmath}
\usepackage{amssymb}
\usepackage{bbold}
\usepackage{color}
\usepackage{hyperref}
\usepackage{amsthm}

\usepackage{tikz}
\usetikzlibrary{quantikz2}

\def\bea{\begin{eqnarray}}
\def\eea{\end{eqnarray}}
\def\<{\langle}
\def\>{\rangle}
\def\tr{\text{tr}}

\def\be{\begin{equation}}
\def\ee{\end{equation}}

\def\non{\nonumber}

\newtheorem*{conjecture*}{Conjecture}

\begin{document}

\title{Maximal Magic for Two-qubit States}

\author{\vspace{0cm} Qiaofeng Liu$^{\, a}$, Ian Low$^{\, a,b}$, Zhewei Yin$^{\, a, b}$}
\thanks{Corresponding author. Email: zheweiyin@gmail.com}
\affiliation{\vspace{0.1cm}
\mbox{$^a$ Department of Physics and Astronomy, Northwestern University, Evanston, IL 60208, USA}\\ 
\mbox{$^b$ High Energy Physics Division, Argonne National Laboratory, Lemont, IL 60439, USA}\\
 \vspace{-0.3cm}}

\begin{abstract}
Magic is a quantum resource essential for universal quantum computation and represents the deviation of quantum states from those that can be simulated efficiently using classical algorithms. Using the Stabilizer R\'enyi Entropy (SRE), we investigate two-qubit states with maximal magic, which are most distinct from classical simulability, and  provide strong numerical evidence that the maximal  second order SRE is $\ln (16/7)\approx 0.827$,  establishing a tighter bound than the prior $\ln(5/2)\approx 0.916$. We identify 480 states saturating the new bound, which turn out to be the fiducial states for the mutually unbiased bases (MUBs) generated by the orbits of the Weyl-Heisenberg (WH) group, and conjecture that  WH-MUBs are the maximal magic states for $n$-qubit, when $n\neq 1$ and 3. 
We also reveal a striking interplay between magic and entanglement: the entanglement of maximal magic states is restricted to two possible values, $1/2$ and $1/\sqrt{2}$, as quantified by  the concurrence; none is maximally entangled.
\end{abstract}

\maketitle

\section{Introduction} 

A major goal in physics and quantum information is to understand the inherent boundary between quantum and classical computing. While entanglement is often cited as the main resource for quantum computing,\footnote{Though we see in cases such as deterministic quantum computation with one qubit (DQC1) \cite{Knill:1998wi} that for certain problems entanglement is not required for quantum speedup.} it is not sufficient for computational speedups due to the Gottesman-Knill theorem \cite{gottesman1998heisenbergrepresentationquantumcomputers,Gottesman:1999tea,Aaronson:2004xuh}, which shows that quantum circuits involving only the Clifford gates and stabilizer states, some of which are maximally entangled, can be simulated on classical computers in polynomial time. A second layer of ``quantumness'' is necessary to characterize the computational advantage of quantum computers, which is the concept of  magic, or non-stabilizerness, introduced by Bravyi and Kitaev in Ref.~\cite{Bravyi:2004isx}. Magic states are believed to be exponentially difficult to simulate using classical algorithms. In turn, universal quantum computation requires the presence of magic states \cite{Bravyi:2004isx}.

Quantifying the amount of magic in a quantum state has become a central question in quantum information and computation \cite{Mari:2012ypq,Veitch_2012,Veitch_2014,Bravyi_2019,PRXQuantum.3.020333,Leone:2021rzd,Turkeshi:2023ctq,Warmuz:2024cft}. Among the several measures the stabilizer R\'enyi entropy (SRE) proposed in Ref.~\cite{Leone:2021rzd},  expressed in terms of the expectation values of Pauli strings, has received considerable attention  because of its ease to compute, as well as being measurable experimentally \cite{Oliviero:2022bqm,Haug:2023ffp}. Although SRE was originally proposed for qubit systems, it has since been generalized to qudits \cite{Wang:2023uog}. SRE is based on the R\'enyi entropy of order $\alpha$ and, in the context of quantum resource theory, has been shown to be a good quantitative measure of the amount of resource in a quantum state for $\alpha\ge 2$ \cite{Leone:2024lfr}.

Beyond quantum computing the concept of magic has  been discussed in diverse physical systems, including quantum many-body physics \cite{Ellison:2020dkj,Oliviero:2022euv,Rattacaso:2023kzm,Gu:2024ure}, quantum field theories \cite{White:2020zoz,Nystrom:2024oeq,Cepollaro:2024sod}, black holes \cite{Hayden:2007cs,Leone:2022afi}, nuclear and astroparticle physics \cite{Robin:2024bdz,Chernyshev:2024pqy},   simulations of quantum gravity \cite{Cepollaro:2024qln}, as well as high energy collider physics \cite{White:2024nuc}.

An important open question in the study of magic and the associated quantum resource is to identify states which are most distinct from the stabilizer states, i.e. quantum states with maximal magic and, therefore, maximal resources. It was pointed out in Refs.~\cite{Wang:2023uog,Cuffaro:2024wet} that, for a $d$-dimensional Hilbert space, an upper bound for SRE of order $\alpha$ is given by,
\be
\label{eq:srebound}
M_\alpha (| \psi \>) \le \frac{1}{1-\alpha}\ln \frac{1+(d-1)(d+1)^{1-\alpha}}{d} \ .
\ee
The bound is saturated if and only if $|\psi\>$ is a Weyl-Heisenberg covariant Symmetric Informationally Complete (WH-SIC) state, which can generate a special, highly symmetric and complete operator basis. Whether SICs exist in all $d$-dimension is an open problem \cite{axioms6030021} and connected to  Hilbert's 12th problem in algebraic number theory \cite{Appleby_2017}. At $d=4$, the bound in Eq.~(\ref{eq:srebound}) is $\ln(5/2)\approx 0.916$ for $\alpha=2$. However,  WH-SIC states do not exist for composite systems at $d=4$ \cite{Cuffaro:2024wet} so the bound is saturated only for a single qudit, but not for two-qubit states. This suggests a more stringent bound than Eq.~(\ref{eq:srebound}) could exist for two-qubit states.

In this work we address the issue of maximum magic for two-qubit states. Not only will we provide a more stringent bound  than Eq.~(\ref{eq:srebound}), we will also enumerate all 480 two-qubit states possessing maximum magic. These maximum magic states turn out to be the fiducial states of mutually unbiased bases (MUBs) \cite{DURT_2010} generated by the orbits of Weyl-Heisenberg (WH) group \cite{1056185,Blanchfield_2014,Feng:2024whz}. The MUB is a foundational notion in quantum physics \cite{Bohr:1928vqa} and important in quantum cryptography and tomography \cite{DURT_2010}. Its existence and constructions are also deeply connected to that of the SICs \cite{WoottersW.K.2006QMaF}.

\section{SRE for Qudits} 

The SRE originally defined for qubits \cite{Leone:2021rzd} has been generalized to qudits in Ref. \cite{Wang:2023uog}. The key is to generalize the Pauli group to the WH group for any dimension. For a $d$-dimensional system, consider an orthonormal computational basis $\{ |k \> \}$, $k = 0, 1, \cdots, d-1$, we can first define the shift operator $X$ and the clock operator $Z$ as
\bea
X| k \> = |k+1 \>,\quad Z | k \> = \omega^k | k \>,
\eea
with $\omega = \exp (2 \pi i /d)$ and $|d\> \equiv |0\>$. We then define the displacement operators
\bea
D_{a_1 a_2} = \omega^{a_1 a_2/2} X^{a_1} Z^{a_2} , \quad a_i = 0,1, \cdots, d-1.
\eea
The WH group  for a single $d$-dimensional qudit is then given by  $\widetilde{\mathcal{W}} (d) = \{\omega^s D_{a_1 a_2} \}$, $s = 0, 1, \cdots, d-1$.
For a single qubit, $\widetilde{\mathcal{W}} (2)$ is the Pauli group consisting of $\{ \pm \mathbb{1}, \pm i \mathbb{1}, \pm X, \pm i X, \pm Y, \pm i Y, \pm Z, \pm i Z \}$, where $\mathbb{1}$ is the identity operator and $\{X , Y , Z \}$ are given by the Pauli matrices $\{ \sigma_1, \sigma_2, \sigma_3 \}$. For a composite system of $n$ qudits, the WH group is given by the tensor product of the WH groups for each single qudit. In the context of magic or non-stabilizerness of the system, we are not interested in overall phases, so we define the quotient group $\mathcal{W} (d) = \widetilde{\mathcal{W}} (d)/\< \omega^s \mathbb{1} \>$ consisting of operators with only $+1$ phases, which is order $d^2$. The SRE of a pure quantum state $|\psi \>$ in a $d$-dimensional Hilbert space is  
\bea
M_{\alpha} (|\psi\>) = \frac{1}{1-\alpha } \ln \sum_{\mathcal{O} \in \mathcal{W} (d) } \frac{1}{d} |\< \psi | \mathcal{O} |\psi \>|^{2 \alpha}.
\eea
We will compute $M_{2} (|\psi\>) = -\ln \Xi_2 (|\psi\>)$, with
\bea
\Xi_2 (|\psi\>) =  \sum_{\mathcal{O} \in \mathcal{W} (d) } \frac{1}{d} |\< \psi | \mathcal{O} |\psi \>|^{4},
\eea
as the measure of magic. For mixed states there is currently no single unique way to extend the definition of the SRE, thus we will focus on pure states within the scope of this work. We do want to point out that for a magic monotone \cite{Leone:2024lfr}, i.e. a suitable measure of magic in the context of quantum resource theories, the measure of magic needs to be nonincreasing under stabilizer protocols, which include introducing classical randomness to the system, e.g. compositing a mixed state from pure states. Therefore, for an extension of the SRE to mixed states that is a magic monotone, such as the convex-roof extension presented in Ref. \cite{Leone:2024lfr}, our determination of maximal magic for pure states will also set the upper bound for mixed states.

The Clifford group is the normalizer of $\widetilde{\mathcal{W}} (d)$ within the full unitary group $U(d)$ of the Hilbert space. For an arbitrary state $|\psi\>$, SRE is invariant under the operation of the Clifford group. The stabilizer states are the eigenstates of maximal Abelian subgroups of $\mathcal{W} (d)$, which have zero magic:  $M_\alpha (| \psi \>) = 0$. The full set of stabilizer states  can be generated by the orbit of the Clifford group for the computational basis. The number of stabilizer states for a single $d$-dimensional qudit, when $d$ is a prime power, is $d(d+1)$ \cite{bandyopadhyay2001newproofexistencemutually}, while the 2-qubit system has 60 stabilizer states \cite{Aaronson:2004xuh}.

\section{Maximal Magic for a Qubit and a Qudit}

Ref. \cite{Cuffaro:2024wet} proves that, for a $d$-dimensional Hilbert space, if there exists a WH covariant symmetric informationally complete positive operator-valued measure (SIC-POVM, or SIC for short), its states uniquely saturate Eq.~(\ref{eq:srebound}). Specifically, for $\alpha = 2$ we have
\bea
\Xi_2 (| \psi \>) \ge \frac{2}{d+1}, \qquad M_2 (| \psi \>) \le \ln \frac{d+1}{2}\ .\label{eq:sicb2}
\eea
The SICs are sets of $d^2$ rank-1 projection operators $\Pi_i  = |\psi_i \> \< \psi_i|$, $i = 1, 2, \cdots, d^2$, such that
\bea
\tr \left( \Pi_i \Pi_j \right) = | \< \psi_i | \psi_j \> |^2 = \left\{ \begin{array}{cc}
    1, & i = j  \\
    \frac{1}{d+1} & i \ne j 
\end{array} \right. .
\eea
An SIC is covariant with respect to group $G$ if each of its normalized fiducial states $|\psi_i \>$ satisfies
\bea
| \< \psi_i | \mathcal{O} | \psi_i \>|^2 = \frac{1}{d+1}, \quad \forall\, \mathcal{O} \ne \mathbb{1} \in G.
\eea
This implies that an SIC  covariant with respect to the WH group $\mathcal{W} (d)$ is an orbit of $\mathcal{W} (d)$.

As an example, consider the system of a single qubit with orthonormal basis $\{ | 0 \>, |1 \> \}$, so that a general state $|\psi \>$ can be represented as a 2-$d$ complex vector $( c_1, c_2 )$ such that $| \psi \> = c_1 |0 \> + c_2 | 1 \>$. The WH group $\mathcal{W} (2)$ is just the Pauli group (with the overall phase modded out): $\{ \mathbb{1}, X, Y, Z \}$. There are  two WH SICs for $d=2$, each containing $d^2 = 4$ states. One example is 
\bea
\!\!\!\!\!\!\!\!&& (  \sqrt{6}+ \sqrt{2}, \left(1-i \right )\sqrt{2 }  ), \ ( \sqrt{3}+2-i ,-1+ \sqrt{3} i   ), \non\\
\!\!\!\!\!\!\!\!&& (   \sqrt{3}-i  ,  \sqrt{3}+ 2+i  ),\ ( -1 + \sqrt{3} i   , \sqrt{3}+ 2-i   ) \, ,\label{eq:2dSIC1}
\eea
up to an overall normalization factor of $1/(2 \sqrt{3+\sqrt{3}} )$. It can be checked that they saturate the bound in Eq.~(\ref{eq:srebound}):  $M_2 = \ln 3/2$; in fact, they  are  the ``$T$-type'' magic states originally defined in Ref. \cite{Bravyi:2004isx}.

For illustration purposes, let us derive the states with maximal magic for a single qubit using $M_2$. Without loss of generality, we can adopt the usual Bloch sphere parameterization for the one qubit state:
\bea
c_1 = \cos ( \theta /2), \quad c_2 = \sin (\theta/2) e^{i \phi},
\eea
with $0\le \theta \le \pi$, $0 \le \phi < 2\pi$. Then we can compute $M_2 = - \ln \Xi_2$ with
\bea
\Xi_2 (| \psi \>) = \frac{8 \sin ^4  \theta  \cos 4 \phi +4 \cos  2 \theta +7 \cos  4 \theta  +53}{64} \, ,  
\eea
for which the global minima can be found analytically.  An example is given by
\bea
\theta = -2\, \text{arccot} \sqrt{2+ \sqrt{3}},\quad \phi = 3\pi/4,\label{eq:2qbs1}
\eea
which generates the first state in Eq.~(\ref{eq:2dSIC1}). The other 7 states are  obtained by the orbit of the Clifford group.

Now let us consider a single $d=4$ qudit. Again without loss of generality we  parameterize the state as
\bea
|\psi\> = c_1\, | 0 \> + c_2\, | 1 \> + c_3\, | 2 \> + c_4\, |  3 \>,
\eea
with
\bea
& c_1 = \sin \theta_1 \sin \theta_2 e^{i\phi_1}, &  c_2 =  \sin \theta_1 \cos \theta_2 e^{i\phi_2},\non\\
& c_3 = \cos \theta_1 \sin \theta_3 e^{i\phi_3}, & c_4 =  \cos \theta_1 \cos \theta_3,\label{eq:4dcp}
\eea
and $\theta_i \in [0, \pi/2]$, $\phi_i \in [0,2 \pi)$. We have fixed the phase of $c_4$ to be $0$, as we are only interested in distinct states that are not related by an overall phase, which does not change magic or entanglement. It is straightforward to work out $M_2 (| \psi \>) = - \ln \Xi_2 (| \psi \>)$ as a function of $\{ \theta_i ,\phi_i\}$, the form of which is included in the supplemental material \cite{supp:4dqdx2}. Because of its complicated form, an analytic derivation of the global minima of $\Xi_2 (| \psi \>)$ turns out to be impractical. One can on the other hand perform a numerical search, which gives a global minimal value of $\min \Xi_2 (| \psi \>) = 2/5$ up to very high precision, agreeing with Eq.~(\ref{eq:sicb2}). It is much less trivial to figure out the maximal magic states, and it is only from our knowledge of WH SICs that we are able to identify them. It is known that there are 256 WH SIC fiducial states for the $d=4$ qudit, which can be partitioned into 16 sets of 16 states, each forming a WH SIC \cite{10.1063/1.1737053}. We have performed a numerical scan of the minima for $\Xi_2$ that identifies all of the 256 states.

\section{Maximal Magic for Two Qubits}

The WH group for a $n$-qubit system is $\mathcal{W} (2)^{\otimes n}$, and it is known that WH SICs only exist for $n=1$ or $3$ \cite{GODSIL2009246}, and the bound given by Eq.~(\ref{eq:srebound}) can not be saturated otherwise. For $d=4$, this means that the bound $M_2 \le \ln (5/2)$ can be saturated by a single qudit, but not a system of two qubits. Intuitively, the SIC treats all the ``directions'' of a Hilbert space on equal footing, which is compatible with the geometry represented by the WH group $\mathcal{W} (4)$. On the other hand, the WH group $\mathcal{W} (2)^{\otimes 2}$ for a tensor product of two Hilbert spaces of $d=2$ realizes the geometry of a different structure, such that its orbit cannot possibly generate an SIC and saturate the bound given by Eq. (\ref{eq:sicb2}).

Nevertheless, it is still straightforward to find out numerically the 2-qubit states with maximal magic, by e.g. the following parametrization of the state $|\psi\>$:
\bea
|\psi\> = c_1\, |00\> + c_2\, |01\> + c_3\, |10\> + c_4\, | 11\>,
\eea
with $|ij\> \equiv |i \> \otimes |j\>$ and $\{c_i \}$ again parameterized by Eq. (\ref{eq:4dcp}). The function $\Xi_2$ is then computed to be
\begin{widetext}
\bea
\Xi_2 (| \psi \>) &=& \frac{3}{32} \sin ^4(2 \theta_1) \sin ^2(2 \theta_2) \sin ^2(2 \theta_3) \left[2 \sin ^2 \phi_3 \sin ^2(\phi_2-\phi_1) +2 \sin ^2 \phi_2 \sin ^2(\phi_3-\phi_1)+2 \sin ^2 \phi_1 \sin ^2(\phi_2-\phi_3)\right.\non\\
&&\left.+1 \right] +2 \sin ^4 \theta_1 \cos ^4 \theta_1 \left[24 \sin ^2 \theta_2 \cos ^2 \theta_2  \sin ^2 \theta_3  \cos ^2 \theta_3  \left(\cos ^2 \phi_3 \cos ^2(\phi_2-\phi_1)+\cos ^2 \phi_2 \cos ^2(\phi_3-\phi_1)\right.\right.\non\\
&&\left. +\cos ^2 \phi_1 \cos ^2(\phi_2-\phi_3)\right) +\cos ^4 \theta_2 \left(\sin ^4 \theta_3 (\cos (4 \phi_2-4 \phi_3)+6) +\cos ^4 \theta_3 (\cos (4 \phi_2)+6)\right)\non\\
&& \left. +\sin ^4 \theta_2 \left(\sin ^4 \theta_3 (\cos (4 \phi_3-4 \phi_1)+6)+\cos ^4 \theta_3 (\cos (4 \phi_1)+6)\right)\right]   \non\\
&&  +2 \sin ^8 \theta_1 \sin ^4 \theta_2 \cos ^4 \theta_2 (\cos (4 \phi_2-4 \phi_1)+6)  +\sin ^8 \theta_1 \sin ^8 \theta_2 +\sin ^8 \theta_1 \cos ^8 \theta_2\non\\
&&+\cos ^8 \theta_1 \left[2 \sin ^4 \theta_3 \cos ^4 \theta_3 (\cos (4 \phi_3)+6)+\sin ^8 \theta_3 +\cos ^8 \theta_3 \right].
\eea
\end{widetext}
A numerical search will give a global minimum of $\min \Xi_2 (|\psi\>)  = 7/16 $ located at
\bea
\theta_{1,2,3} = \frac{\pi}{4}, \quad \phi_{1,2,3} = \frac{\pi}{2}\label{eq:2qsol1}
\eea
within our numerical precision. This corresponds to the state
\bea
( i,i,i,1)/2,\label{eq:mlex}
\eea
for which we analytically confirms that $\Xi_2$ is indeed exactly $7/16$. We checked that the gradient  $\nabla \Xi_2$ with respect to $\{\theta_i, \phi_i \}$ at the position of Eq.~(\ref{eq:2qsol1}) gives exactly $0$, and the Hessian matrix $\nabla \otimes \nabla \Xi_2$ at this position is positive definite, thus confirming that Eq.~(\ref{eq:mlex}) corresponds to an isolated local minimum of $\Xi_2$ without a flat direction. Namely the minimum is not continuously connected to other  minima (except through an overall phase). We are thus confident that the above indeed is a location for the global minimum of $\Xi_2$, which is $7/16$. The maximal magic achievable for a 2-qubit system is then
\bea
\max \left[ M_{2} (|\psi \>) \right] = \ln \frac{16}{7} \approx 0.827.
\eea

Similar to what we have done for a single qubit, we can start with the state specified by Eq.~(\ref{eq:mlex}) and find all other maximum magic states via the orbit of the Clifford group,  which generates  a total of 480 distinct states. Examples include
\bea
\left( 0, i, -i, 1+i \right)/2, \ \left( 1+i , 1+i , 1-i , 3+i \right)/4.\label{eq:conhe}
\eea
As a sanity check we also performed a brute-force numerical scan of the parameter space to find exactly 480 states giving rise to the global minima of $\Xi_2$.

\begin{figure}[tbp]
    \centering
    \begin{quantikz}
    & \gate{T} & \ctrl{1} &                  & \\
    & \gate{T} & \targ{}  & \gate{T} & 
    \end{quantikz}
    \caption{The sequence of gates that turns the stabilizer state $(1,1,1,1)/2$ into the maximal magic state $(1,i,i,i)/2$.}
    \label{fig:cir}
\end{figure}
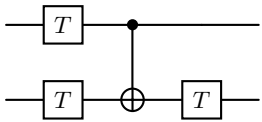

It is well-known that the Clifford gates, augmented by the single-qubit $T$ gate, are universal for quantum computation, which means any quantum logic gates can be well-approximated by a finite sequence of Clifford and $T$ gates. It turns out that for  2-qubit states, the maximal magic states can be \textit{exactly} generated from a stabilizer state by going through a finite number of these gates. This can be easily seen from the fact that $(1,1,1,1)/2$ is a stabilizer state while $(1,i,i,i)/2$ is a maximal magic state, and the latter can be generated by the former going through the following simple sequence:  $(\mathbb{1}\otimes T) \cdot \text{CNOT}_{12} \cdot (T \otimes T)$, as shown in Fig.~\ref{fig:cir}.

\section{Maximal Magic and WH MUBs}

Although the maximal magic states for two qubits are not  WH SICs, they turn out to be another group of special states known as the fiducial states for WH MUBs \cite{1056185,Blanchfield_2014,Feng:2024whz}. Consider two orthonormal bases $|\psi_i \>$ and $|\chi_i\>$ in a $d$-dimensional Hilbert space; they are mutually unbiased (MU) if $|\< \psi_i | \chi_j \>|^2 = 1/d$ for any $i,j = 0,1, \cdots, d-1$. It is known that  the  number of bases satisfying the above criterion is at most $d+1$, which can be saturated when $d$ is a prime-power \cite{WOOTTERS1989363}. Namely, for $d=4$ there are at most 5 MUBs. Actually, for a single qudit of a prime-power dimension $d$, the $d(d+1)$ stabilizer states can be partitioned into $d+1$ MUBs \cite{bandyopadhyay2001newproofexistencemutually}.

A state $|\psi\>$ is a WH MUB fiducial state if the orbit $\mathcal{W} (d) |\psi \>$ generates a set of $d^2$ non-equivalent states, which can be partitioned into $d$ MUBs. From the definition of MUB, one can easily compute the magic of a WH MUB fiducial state to be
\bea
M_\alpha (|\psi\>) = \frac{1}{1-\alpha} \ln \frac{1 + (d-1) d^{1-\alpha}}{d},\label{eq:srewhmub}
\eea
which is clearly below the bound given by WH SICs in Eq.~(\ref{eq:srebound}). For $d=4$ and $\alpha=2$, the above gives $\ln 16/7$ which agrees with the two-qubit maximum magic   we found through the numerical search.

One can check that for a stabilizer state $|\psi\>$ the orbit $\mathcal{W} (2)^{\otimes 2} |\psi \>$ leads to only 4 unique states, which form an orthonormal basis; as the WH group is a subgroup of the Clifford group, this basis consists of stabilizer states. The 60 stabilizer states can be partitioned into 15 bases, corresponding to 15 different orbits of $\mathcal{W} (2)^{\otimes 2} $, which can be further grouped into 3 families of 5 MUBs. On the other hand, the 480 states that maximize magic are all WH MUB fiducial states: They can be partitioned into 30 families, each corresponding to 4 MUBs that form an orbit of $\mathcal{W} (2)^{\otimes 2} $. It turns out that each of the 30 orbits of the WH group for the maximal magic states can be augmented by one of the 15 orbits of $\mathcal{W} (2)^{\otimes 2} $ for the stabilizer states to form 5 MUBs, which is the maximum number of MUBs for $d=4$. Specifically, each of these 15 orbits for stabilizer states is in this manner uniquely associated with 2 of those 30 orbits for MUB fiducial states. A full list of stabilizer states and maximal magic states, organized according to the orbits of $\mathcal{W} (2)^{\otimes 2} $ and the MUB associations, is presented in \cite{supp:states}.  One example for such a family of 5 MUBs consists of one basis formed by the stabilizer states
\bea
(1,0,0,0),(0,0,1,0),(0,1,0,0),(0,0,0,1),\label{eq:sobe}
\eea
and 4 MUBs formed by maximal magic states and corresponding to an orbit of $\mathcal{W} (2)^{\otimes 2} $ (up to a normalization factor of $1/2$):
\bea
\!\!\!\!\!\!&&(1,-1,-1,i),(1,-1,1,-i),(1,1,-1,-i),(1,1,1,i);\non\\
\!\!\!\!\!\!&&(-1,i,1,-1),(i,1,i,-i),(-1,-i,1,1),(i,-1,i,i);\non\\
\!\!\!\!\!\!&&(-1,1,i,-1),(-1,1,-i,1),(i,i,1,-i),(i,i,-1,i);\non\\
\!\!\!\!\!\!&&(i,-1,-1,1),(1,i,-i,i),(1,-i,i,i),(i,1,1,1),\label{eq:mobe}
\eea
where each line of the above corresponds to a MUB.

As pointed out by Ref. \cite{Feng:2024whz}, WH MUB fiducial states in general realize a local maximum for magic. In this work we see that for two-qubit states they generate the global maximum for magic. We thus propose the following conjecture:

\begin{conjecture*}
For a system with no WH SIC, the maximal SRE $M_{\alpha}$, $\alpha \ge 2$,\footnote{$M_\alpha$ for $\alpha < 2$ may not act as a suitable measure of magic in contexts such as magic-state resource theory \cite{Haug:2023hcs,Leone:2024lfr}. We note that numerical evidence suggests that for 2 qubits, instead of the WH MUB fiducial states, the state $( 0,-1+ i\sqrt{3},\sqrt{3}-i , 2 )/(2\sqrt{3})$ actually maximizes $M_\alpha$ for $\alpha < 1.64 \cdots$, which includes the interesting case of $\alpha = 1/2$ corresponding to the ``stabilizer norm'' \cite{Howard:2017maw,Hahn:2021jya,Dai:2022zgo} utilized in Ref. \cite{Feng:2024whz}.} is uniquely saturated by the WH MUB fiducial states on the condition that they exist.  Specifically, for $n$-qubit systems with $n\ne 1,3$, the maximal SRE $M_{\alpha}$, $\alpha \ge 2$, is given by
Eq.~(\ref{eq:srewhmub}), which is uniquely saturated by the WH MUB fiducial states.
\end{conjecture*}

Notice that in the large $d$ limit, both the bound given by Eq. (\ref{eq:srebound}) and by Eq. (\ref{eq:srewhmub}) approaches $\ln d$, at the leading order of $1/d$. It is known that the average magic for random states approaches $\ln d$ for large $d$ as well \cite{Leone:2021rzd,Chen:2022yza,Chen:2023rzi,Szombathy:2025euv,Bittel:2025yhq}, indicating that they contain almost maximal non-stabilizerness.

Lastly, the entanglement between the two qubits for stabilizer states is either minimal or maximal. The amount of entanglement can be quantified by e.g. the concurrence \cite{Hill:1997pfa,Wootters:1997id}, defined for a pure state as
\bea
\Delta (| \psi \>) = 2|c_1 c_4 - c_2 c_3|.
\eea
As $\mathcal{W} (2)^{\otimes 2} $ consists of operations that factorize for the two qubits, the operation of the WH group do not change the amount of entanglement, thus an orbit of $\mathcal{W} (2)^{\otimes 2} $ consists of states with the same concurrence. For stabilizer states, 9 out of the 15 orbits of $\mathcal{W} (2)^{\otimes 2} $  consist of unentangled product states, $\Delta = 0 $, while the rest are in maximally entangled states with $\Delta = 1$. On the other hand, the 30 orbits of $\mathcal{W} (2)^{\otimes 2} $ for MUB fiducial states, while maximizing magic, do not minimize or maximize entanglement. They realize a concurrence of either $1/\sqrt{2}$ or $1/2$. Specifically, the orbit of the WH group for MUB fiducial states realizes $\Delta = 1/\sqrt{2}$ if it can be augmented by an unentangled  stabilizer orbit of $\mathcal{W} (2)^{\otimes 2} $ to form a family of 5 MUBs, while the ones associated with the maximally entangled stabilizer orbits of $\mathcal{W} (2)^{\otimes 2} $ correspond to $\Delta = 1/2$. For example, states in Eqs.~(\ref{eq:sobe}) and (\ref{eq:mobe}) have $\Delta = 0$ and $\Delta = 1/\sqrt{2}$, respectively, while those in Eq. (\ref{eq:conhe}) has $\Delta = 1/2$.

\section{Conclusions}

In this work we use the second order SRE to study maximum magic in two-qubit states and discover numerically an upper bound of $\ln(16/7)$. We also establish 480 states saturating the maximal magic and identify them to be the fiducial states for the mutually unbiased bases generated by the orbit of the Weyl-Heisenberg group. The previous, weaker bound of $\ln(5/2)$ is based on the maximal magic carried by the Weyl-Heisenberg covariant Symmetric Informationally Complete states, and for $n$-qubit systems they only exist in the one- and  three-qubit cases. We also conjecture that the MUBs are the maximal magic states when the WH-SICs do not exist and provide an analytic expression for the SRE of the MUBs. In addition, the conjecture will apply to all $n$-qubit states, for $n\neq 1,3$. It turns out there is an interesting interplay between magic and entanglement -- the concurrence of maximal magic states falls into only two values, $1/2$ and $1/\sqrt{2}$, and none is maximally entangled.

Characterizing states with maximal magic is an important topic in magic resource theory \cite{Veitch_2014,RevModPhys.91.025001,PhysRevLett.123.020401,PRXQuantum.3.020333} and has major implications in many areas of quantum computation and quantum physics, including quantum cryptography \cite{PhysRevLett.132.210602},  fidelity estimation \cite{PhysRevA.107.022429}, entanglement dynamics \cite{gu2024magicinducedcomputationalseparationentanglement}, and quantum chaos \cite{Garcia_2023}, just to name a few. Our work represents a crucial step toward classifying all states with maximum magic and opens up new venues for exploring the connections between MUBs and SICs, two foundational concepts in quantum information. As the SRE is far from the only measure of magic, it is important to investigate whether the maximum SRE states also saturate other such measures. Last but not least, it will be interesting to study the significance of the maximal magic states in the various physical systems where magic plays an important role. 

\section{Acknowledgements}
We thank Zi-Wen Liu for valuable comments on the manuscript.
This work is supported in part by the U.S.
Department of Energy, Office of High Energy Physics,
under contract DE-AC02-06CH11357 at Argonne, and by the U.S. Department of Energy, Office of Nuclear
Physics, under grant DE-SC0023522 at Northwestern.

\bibliography{ref}

\end{document}